\documentclass[11pt,twoside]{article}

\usepackage{asp2006}
\usepackage{epsf}
\usepackage{epsfig}
\usepackage{lscape}
\usepackage{natbib}

\begin{document}
\title{Metal Abundances of Subdwarf B Stars from SPY - a Pattern Emerges}
\author{S. Geier$^1$, U. Heber$^1$, and R. Napiwotzki$^2$}  
\affil{$^1$ Dr.--Remeis--Sternwarte, Institute for Astronomy, University Erlangen-Nuremberg, Sternwartstr. 7, 96049 Bamberg, Germany}
\affil{$^2$ Centre of Astrophysics Research, University of Hertfordshire, College Lane, Hatfield AL10 9AB, UK}


\begin{abstract} 
The formation of sdBs is still puzzling,  as is the chemical
composition of their atmospheres. While helium and other light
elements are depleted relative to solar values, heavy elements are
highly enriched. Diffusion processes in the hot, radiative atmosphere
of these stars are the most likely explanation. Although several
attempts were made, it was not yet possible to model all the observed
features of sdB atmospheres. A setback of most prior studies was the
small sample size. We present a detailed abundance analysis of 68
sdBs. From high resolution spectra obtained with the VLT/UVES
instrument in the course of the ESO Supernova Progenitor Survey (SPY)
we measured elemental abundances of up to 24 different ions per
star. A general trend of enrichment was found with increasing
temperature for most of the heavier elements. The lighter elements
like carbon, oxygen and nitrogen are depleted irrespective of the
temperature. Although there is considerable scatter from one star to
another, the general abundance patterns in most sdBs are similar. An
interplay between gravitational settling, radiative levitation and
weak winds is most likely responsible. About $6\%$ of the analysed
stars show an enrichment in carbon and helium which cannot be
explained in the framework of diffusion alone. Nuclear processed
material must have been transported to the surface. The late 
hot-flasher scenario may provide a possible explanation for this effect.
\end{abstract}


\section{Introduction}   

Hot subdwarf stars are very important for the study of stellar
atmospheres. The chemical peculiarities of their atmospheres are still
puzzling although research has been done in this field for more than
four decades. \citet{pii2_sargent} discovered the helium deficiency of sdB
stars for the first time. Peculiar metal abundances were
reported. While some metals showed solar abundances, others were
depleted or even enriched. Theoretical diffusion models yielded only
little success. Radiative levitation and mass loss caused by stellar
winds counteract the gravitational settling \citep{pii2_fontaine,
pii2_unglaub}. Since the discovery of pulsating sdB stars, the metal content
in their atmospheres became important for asteroseismology as
well. Several abundance studies of sdBs have been undertaken so
far. Most of the data was taken in the UV with different instruments
\citep[IUE, FUSE, HST/STIS, e.g.][]{pii2_otoole}. Some optical spectra at
high resolution were taken with large telescopes
(e.g. Keck/HIRES). The samples analysed so far consisted of only a few stars.

\section{Observations and Data Analysis}

Sixty eight sdBs were observed in the course of the SPY project with
the 
high-resolution spectrograph UVES at the ESO\,VLT. In order to derive the
metal abundances we compared the observed spectra with 
rotationally-broadened, synthetic line profiles using an automatic analysis
pipeline.  For a standard set of 69 metal lines from 24 different ions
an LTE model spectrum with appropriate atmospheric parameters
\citep{pii2_lisker} was automatically generated with LINFOR and fitted
using the FITSB2 routine. Ionization equilibria have been checked to
be consistent in general to within the error limits, except for a few
notorious cases.

\section{Metal Abundances}

We searched for trends of the metal abundances with atmospheric
parameters \citep[see][]{pii2_edelmann} and found correlations only with
the effective temperature $T_{\rm eff}$ (see Fig. 1). The observed
carbon abundances are subsolar and show a large scatter from star to
star. Four exceptional sdBs show nearly solar to supersolar abundances
up to $+1.0\,{\rm dex}$ (see next section). The nitrogen and oxygen
abundances range from $-1.5\,{\rm dex}$ to solar. Neon and magnesium
are also depleted. A trend with temperature is present in the
aluminium and phosphorus abundances. Aluminium is enriched by
$1.5\,{\rm dex}$, phosphorus by $1.0\,{\rm dex}$. In contrast to this,
the silicon and sulfur abundances show a large scatter of $2.5\,{\rm
dex}$ and no trend with temperature. Argon is enriched to $+2.1\,{\rm
dex}$ and shows a clear trend with temperature. Potassium is
identified in an sdB for the first time. It is strongly enriched up to
$+2.9\,{\rm dex}$ and correlated with $T_{\rm eff}$. Doubly ionized
calcium appears at higher temperatures with abundances of up to
$+2.0\,{\rm dex}$. Scandium, titanium, vanadium, and chromium are
highly supersolar ($+2.0$ to $+4.0\,{\rm dex}$). Scandium and chromium
show a clear trend with temperature, which is less pronounced for
titanium and vanadium. The iron abundance shows no trend all over the
parameter space and remains nearly solar. Cobalt could be measured in
five stars only and upper limits are given for all others.

\section{Abundance Pattern in sdB Atmospheres and Diffusion}

The metal abundances of the programme stars show a universal pattern,
which could not been seen before. While the light elements carbon,
nitrogen, oxygen, neon, magnesium as well as iron are not correlated
with temperature, all heavier elements from aluminium to chromium are
more abundant the hotter the star gets. Diffusion timescales are much
shorter than the lifetime on the EHB hence an equilibrium abundance
between radiative levitation, gravitational settling and a weak
stellar wind should be reached. At higher temperatures, where
radiative levitation plays a more important role, this equilibrium
should be shifted to higher abundances of those heavy elements, which
are not abundant in the sun. The lower the elemental abundance, the
higher the UV photon flux and therefore the radiation pressure. An
increase in the abundance leads to a decrease of the photon flux and a
saturation effect \citep{pii2_vauclair}. This is exactly what can be seen
in the data. While the equilibrium abundance does not change in case
of the light elements (except helium) within the sdB temperature
range, all elements heavier than magnesium are enriched. Since the
primordial iron abundance is much higher than the abundances of the
other heavy elements, saturation is more likely and seems to be the
reason for the constant iron abundances. Although the general pattern
may be qualitatively explained in this way, some questions remain
open. All observed trends are superimposed by a scatter from star to
star. The abundance difference between stars with similar temperature
can be as high as $2.0\,{\rm dex}$ for some elements (e.g. silicon or
sulfur). On the other hand there is little scatter for other elements
(e.g. nitrogen or iron).

Four sdBs show a strong enrichment in carbon from nearly solar to ten
times solar. Their helium abundance is high compared to the rest of
the sample. They are situated at the hot end of the EHB, but this
region is not exclusively occupied by these carbon-rich
sdBs. Other subdwarfs with very similar atmospheric parameters don't show
any carbon enrichment.

If diffusion in the stellar atmosphere is the only reason for the
observed abundance patterns, why do we see two different carbon
abundances at the same atmospheric parameters? A mechanism is needed
which constantly transports helium and carbon in the atmosphere and
therefore counteracts diffusion at least to some extent. Weak
photospheric convection may be this mechanism. Non-canonical formation
scenarios like WD merger or late hot flasher \citep{pii2_lanz} may provide
the initial enrichment of helium and carbon necessary to form a small
convection zone \citep{pii2_groth}. Carbon-rich sdBs may therefore be
formed in a similar way as He-sdB/sdOs \citep{pii2_stroer}.

\begin{figure}[tp!]
\centering
 \includegraphics[width=9.7cm,bb=18 144 592 690]{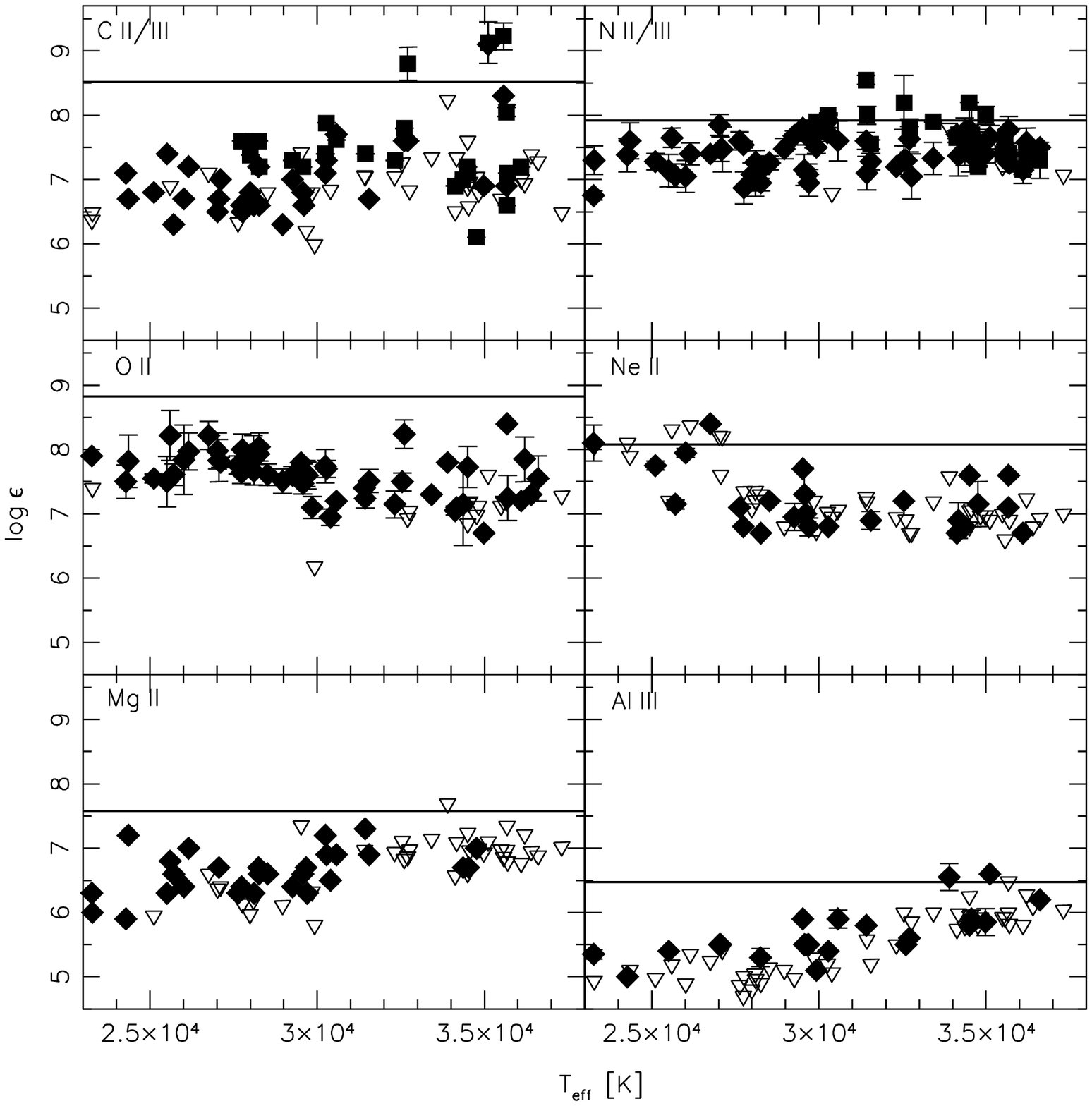}
 \includegraphics[width=9.7cm,bb=18 158 592 690]{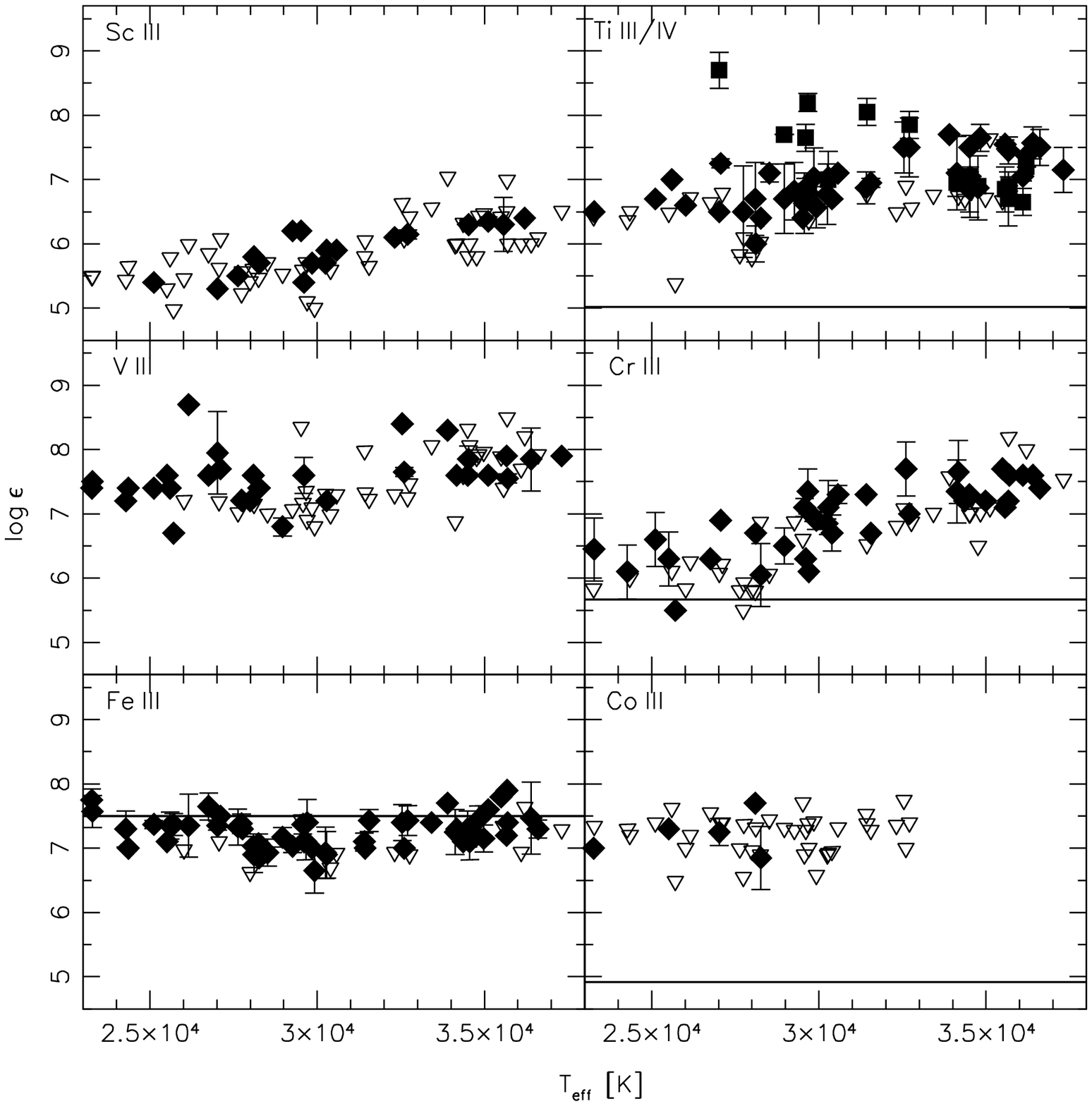}
 \caption{The abundances of selected elements are plotted against the
 effective temperature. If two ionization stages are present, the
 lower one is marked by diamonds, the higher one by rectangles. Open
 triangles mark upper limits. Solar abundances \citep{pii2_grevesse} are
 drawn as horizontal lines. The solar abundances of scandium
 ($\log\,\epsilon_{\rm Sc}=3.17$) and vanadium ($\log\,\epsilon_{\rm
 V}=4.00$) are below the plotted range.}
\end{figure}

\end{document}